\documentclass[conference]{IEEEtran}
\IEEEoverridecommandlockouts
\usepackage{mathtools}
\usepackage{graphicx}
\usepackage{epsfig}
\usepackage{caption}
\usepackage{subcaption}
\usepackage[none]{hyphenat}
\usepackage{multirow}
\usepackage{amsmath,etoolbox}
\usepackage{algorithm}
\usepackage[noend]{algpseudocode}
\usepackage{amssymb}
\usepackage[hyphens]{url}
\usepackage{cite}

\begin{document}

\title{
Cross-layer Band Selection and Routing Design for Diverse Band-aware DSA Networks
}

\author{\IEEEauthorblockN{Pratheek S. Upadhyaya, Vijay K. Shah, and Jeffrey H. Reed \\
Wireless@VT, Bradley Department of Electrical and Computer Engineering \\ Virginia Tech, Blacksburg, VA, USA, 24061. \\
\{pratheek, vijays, reedjh\}@vt.edu
\thanks{This work is partially supported by NSF grant NeTS-1564148. This paper is accepted for publication in IEEE GLOBECOM 2020.}
}}

\maketitle

\begin{abstract} 
As several new spectrum bands are opening up for shared use, a new paradigm of \textit{Diverse Band-aware Dynamic Spectrum Access} (d-DSA) has emerged. d-DSA equips a secondary device with software defined radios (SDRs) and utilize whitespaces (or idle channels) in \textit{multiple bands}, including but not limited to TV, LTE, Citizen Broadband Radio Service (CBRS), unlicensed ISM. In this paper, we propose a decentralized, online multi-agent reinforcement learning based cross-layer BAnd selection and Routing Design (BARD) for such d-DSA networks. BARD not only harnesses whitespaces in multiple spectrum bands, but also accounts for unique electro-magnetic characteristics of those bands to maximize the desired quality of service (QoS) requirements of heterogeneous message packets; while also ensuring no harmful interference to the primary users in the utilized band. Our extensive experiments demonstrate that BARD outperforms the baseline dDSAaR algorithm in terms of message delivery ratio, however, at a relatively higher network latency, for varying number of primary and secondary users. Furthermore, BARD greatly outperforms its single-band DSA variants in terms of both the metrics in all considered scenarios.
\end{abstract}

\maketitle
\section{Introduction} \label{intro}
The radio frequency spectrum that can be used for wireless communications is a finite and extremely valuable resource~\cite{NSF-SII, NSF-SWIFT}. 
There has been a dramatic growth of new wireless communication and application technologies, such as, intelligent transportation systems, unmanned aircraft systems, telehealth, high-definition holographic videos and massive Internet of Things. These and other applications, including 5G and beyond wireless communication systems continue to drive the growth in the demand for radio spectrum and thus, further worsen the spectrum scarcity problem~\cite{presidential-memorandum}. 

This can mainly be attributed to the static nature of current spectrum policy by the Federal Communications Commission (FCC) in United States, where most spectrum bands are licensed and authorized to licensed users (or primary users, PUs) for some dedicated applications. This results in severely underutilized licensed bands (e.g., TV, LTE) whereas overcrowded unlicensed bands (ISM and U-NII band) due to increasing wireless and mobile devices in these bands~\cite{jagannath2018design}. 

In order to remedy this and circumvent the spectrum scarcity issue, the paradigm of Dynamic Spectrum Access (DSA) was proposed in early 2000's~\cite{arslan2007cognitive}. Using DSA, a software defined radio (SDR) user (or secondary user, SU) can opportunistically use the underutilized or unused channels (called whitespaces) in a certain licensed band on the condition of non-interference to the PUs. DSA can improve spectrum utilization dramatically and provide broadband wireless services by using spectrum sharing schemes. However, existing works in DSA literature either consider (i) an unspecified licensed band~\cite{tang2016joint} - a licensed primary band is represented as a frequency interval $[f_{min}, f_{max}]$, which comprises of several non-overlapping orthogonal channels, each having a prespecified transmission range and channel bandwidth, or (ii) \textit{one individual licensed band}, mostly TV band~\cite{jagannath2018design, chowdhury2011crp}.

The current DSA approaches of isolated management of an individual shared band (or adjacent band) limits the potential for efficient sharing with multiple types of PUs as new bands are shared. For instance, new bands such as, $1300 - 1350$ MHz, $1400 - 1427$ MHz, $23.6 - 24$ GHz, $50.2 - 50.4$ GHz, are being considered for shared use as highlighted in current NSF SWIFT proposal call~\cite{NSF-SWIFT}. Thus, the limitations of current DSA approaches point to the need for a new and efficient approach that must consider the flexible use of bands, ranging from TV, LTE, to new spectrum bands. Recently a new paradigm, termed, \textit{Diverse Band-aware DSA} (d-DSA) was proposed by Shah et al.~\cite{shah2017designing, shah2019diverse}, where a d-DSA device (or SU) is equipped with SDRs and utilize whitespaces (or idle channels) in \textit{multiple} spectrum bands, be it, TV, LTE, CBRS or Radar, and unlicensed ISM band. This can be easily extended to new bands that are opening up for shared use.

Unlike conventional (single-band) DSA neworks, since d-DSA networks consider ``diverse spectrum bands'', it requires a fundamental rethinking in the way we design networking solutions. This is mainly because various bands (e.g., TV and CBRS) have unique electro-magnetic (EM) properties, e.g., transmission range, channel bandwidth, PU activities etc. TV band, operating at low frequency bands ($54 - 698$ MHz) offers a very high transmission range (tens of Kms) and low bandwidth (6 - 9 MHz), whereas CBRS ($3.5$ GHz) offers a very high bandwidth (upto $40$ MHz) but suffers in terms of transmission range (few hundred meters). 
Furthermore, it is intuitive that the number of interfering PUs would likely be larger in high-range TV band than that of low-range CBRS band. These unique EM characteristics of diverse bands and the notion of d-DSA paradigm makes the design of cross-layer band selection and routing in d-DSA networks -- a non-intuitive and challenging  problem, which is the focus of our work.

In this paper, we propose a decentralized, online multi-agent reinforcement learning based cross-layer BAnd selection and Routing Design (BARD) for such d-DSA networks. BARD harnesses whitespaces (or idle channels) in multiple  spectrum bands, while taking into account -- the unique EM characteristics and the PU activities in diverse spectrum bands -- to maximize the desired quality of service (QoS) requirements, while also guaranteeing non-interference to the PUs in the utilized band. 
Our experiments demonstrate that BARD outperforms existing centralized dDSAaR protocol~\cite{shah2019x}, in terms of message delivery ratio with relatively higher network latency, for varying number of PUs and SUs. 

The paper is organized as follows: Section \ref{sec:related_works} reviews the existing traditional and RL based routing protocols in DSA and d-DSA networks. Section \ref{sec:network_model} presents the network model, followed by proposed BARD algorithm in Section \ref{sec:bard_algorithm}. In Section \ref{sec:results}, we validate the BARD algorithm against existing routing approaches via extensive simulation experiments. Section \ref{sec:conclusions} concludes the paper.

\section{Related Work} \label{sec:related_works}

Routing in DSA (or Cognitive Radio Networks, CRNs) has been fairly well investigated in the past~\cite{jagannath2018design, tang2016joint, chowdhury2011crp, elrhareg2019routing, liurl2020, rlCRHasan, rlbandchannel2019, rlroutingsurvey}.
The main idea of existing DSA routing protocols is to select ``best available channel'' (loosely referred to as, spectrum selection) and ``an optimal next-hop node'' while accounting for several parameters (like channel identification, switching delay, PU modeling etc.). The authors in \cite{chowdhury2011crp} proposed a routing protocol that selects suitable route and channel between a node-pair, while ensuring protection to PUs. 
Jagannath et al~\cite{jagannath2018design} proposed a joint route, channel and power control algorithm to maximize the utilization of available resources and ensure delivery of packets within the deadline constraints. 
Refer to a recent survey~\cite{elrhareg2019routing} for an overview on DSA routing protocols.  Existing works on DSA routing either assume a DSA network model with an unspecified primary band~\cite{tang2016joint} or one restricted primary band -- mostly TV~\cite{chowdhury2011crp, jagannath2018design}. Thus, these DSA routing protocols are inapplicable to d-DSA networks as they lack the capability to utilize the whitespaces in multiple bands, and do not account for unique EM characteristics of various bands. 

Reinforcement learning (RL) is being widely used for routing in DSA networks (and wireless networks) since it is suited for solving optimization problems in a distributed, decentralized setting. ~\cite{rlCRHasan} used RL to exploit under-utilized licensed spectrum to improve network performance in a cognitive radio setting. 
In~\cite{liurl2020}, Liu et al. propose a hybrid RL approach 
to reduce the sample size and accelerate the performance of model free routing algorithm in DSA networks. \cite{DeepRLMultiDSA} considered a distributed RL based approach for DSA networks to maximize the network utility without any control messages between users. 
A dynamic band and channel selection algorithm was proposed in \cite{rlbandchannel2019} to utilize whitespaces in spectrum and mitigate the effect of PU's interference. While the authors did consider multiple bands, the work was limited to High Frequency (HF), Very High Frequency (VHF) and Ultra High Frequency (UHF) bands with a main focus on channel selection, and largely ignore the distinct EM characteristics of various bands, such as, channel bandwidth, transmission range etc. Refer to ~\cite{rlroutingsurvey} for a detailed overview of RL based routing approaches for DSA networks.

Recently Shah et al. proposed the notion of d-DSA paradigm, and presented a centralized dynamic programming based routing protocol for static d-DSA networks~\cite{shah2017designing, shah2018designing} whereas georouting principle based routing protocol for mobile d-DSA delay-tolerant networks~\cite{shah2020diverse}. In \cite{shah2019x}, the authors proposed a centralized diverse DSA aware routing (dDSAaR) protocol, that constructs a unified Space-Time-Band (STB) graph and computes the entire communication routes for each source-destination node pair in the d-DSA network. 


\section{d-DSA Network Model} \label{sec:network_model}
\begin{figure}[h!]
\vspace{-0.1in}
\centering
  \includegraphics[scale=0.55]{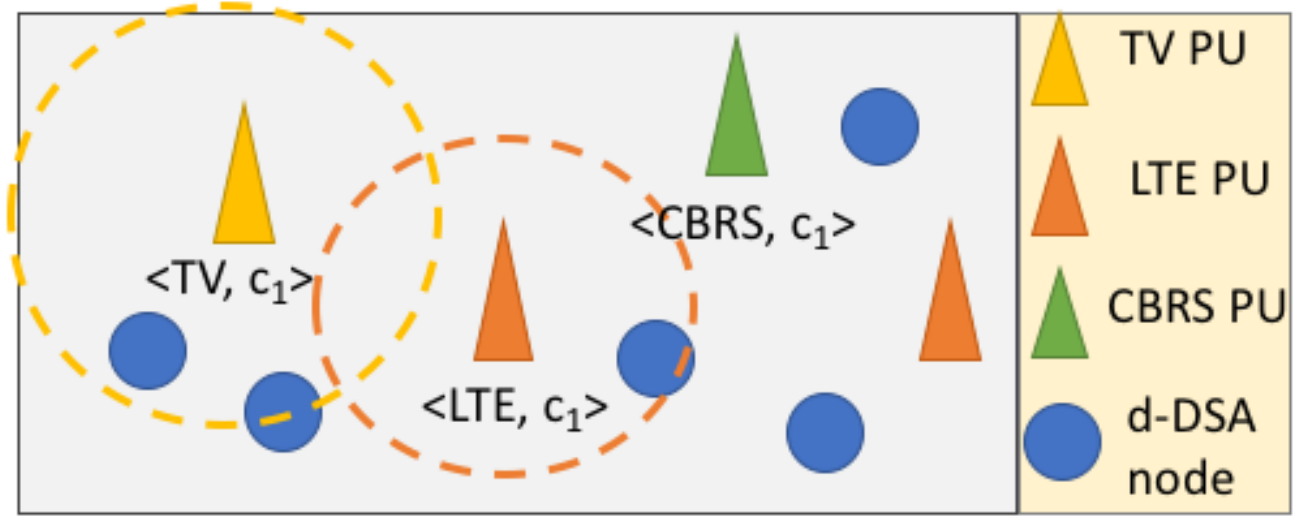}
  \caption{d-DSA network model. A certain PU may operate on a certain channel (e.g., $c_1$) in a prespecified band, say TV. 
  }\label{fig:system_model}
  \vspace{-0.05in}
\end{figure}

As shown in Fig. \ref{fig:system_model}, we consider a d-DSA network, that constitutes several (i) secondary users (SUs), i.e., d-DSA nodes, and (ii) primary users (PUs) that may operate in one of the spectrum band from the band set \{TV, LTE, ISM\footnote{Note unlike other bands, ISM is an unlicensed band and all the devices -- ISM access points (APs) or d-DSA nodes have equal access to the band.}, CBRS\}. 

We model our d-DSA network as a directed graph $G = \{U \cup {P}, \mathcal{B}, \mathcal{E}\}$, where $U$ and $P$ denote the set of d-DSA nodes (SUs) and PUs in the network respectively, $\mathcal{B}  = \{1, 2, \dots ,|B|\}$ is the set of spectrum bands and $\mathcal{E}$ is the set of edges representing the communication links among d-DSA (or SU) nodes. 
A communication link $e_{i,j}^b \in \mathcal{E}$ exists between node $i$ and $j$ over band $b$ only (i) if there is a free common channel available in band $b$ and (ii) if the geographical distance between them $d_{ij}$ is within the communication range $\Delta_b$ (i.e., $d_{ij} \leq \Delta_b$).
Each band $b \in \mathcal{B}$ consists of a predefined set of available channels $C^b = \{c^b_1, c^b_2, \dots, c^b_{|C|}\}$. 
Note that a certain channel $c^b_i$ in a band $b$ may be different from a channel $c^{b'}_i$ in another band $b^{'}$ in terms of channel bandwidth, transmission range and maximum allowable transmit power. The channel access is modeled as a Listen-Before Talk scheme, known as interweaving scheme where a SU can access a channel in the utilized band only if it is available~\cite{Distdynamicchsense}.


\textit{Message heterogeneity.}
Although messages can be of arbitrary sizes, we consider a prespecified $\mathcal{M}$ discrete bins of message sizes, say, $\{1, \dots, q, \dots ,M\}$. Note that a certain message $m$ with size $m_q$ is composed of $n = \frac{m_q}{l}$ packets where $l$ is the packet size. Thus for successful delivery of message $m$, all $n$ packets (corresponding to message $m$ with size $m_q$) must be delivered to the intended destination node.


\textit{Primary User Activity.}
The PU's activity, i.e., presence or absence of the PU signal, (or wireless channel $c^b_i$ in the operating band $b \in \mathcal{B}$) can be modeled as continuous-time, alternating ON/OFF Markov Renewal Process (MRP)~\cite{chowdhury2011crp}. 
Such a model captures the time period in which the channel can be utilized by SUs without causing any harmful interference to PU nodes. The PU activity in each channel is either ON or OFF state, representing busy (presence of PU's activity) or idle (no PU's activity), respectively.

\section{ BARD Algorithm} \label{sec:bard_algorithm}

This section presents the online, decentralized cross-layer band selection and routing Design (BARD) algorithm. 

BARD can be modeled as a multi-agent reinforcement learning (RL) algorithm, which can be formalized using markov decision process (MDP)~\cite{zhang2018fully}. Each d-DSA node acts as an RL agent and learns its joint band selection and routing policy by observing the local network state and communicating with neighbor nodes. Next, we will present the definitions of each element in RL for a single agent.

\textbf{State space.} Let $p$ be the current packet to be sent by the agent. The state space of the agent is denoted as $S:<y_p, m_q, E, E^{'}>$, where $y_p$ is the destination of current packet $p$, $m_q$ is the message size corresponding to packet $p$. $E$ is the spectrum band and channel related information at the agent (i.e., channels in use by PUs and other SUs in each considered band). $E^{'}$ represents similar spectrum band (and channels) and queue related information at neighboring nodes.

\textbf{Action space.} The action space of the agent is defined as $A: <j, b>$, where $j$ is a neighboring d-DSA node of the agent and $b$ represents the band over which a link exists between the agent and d-DSA node $j$. For each packet $p$ at the head of the queue at time stamp $t$, the agent observes the current state $s_t \in S$ and selects an action $a_t \in A$, and the current packet is delivered over a certain band $b$ to an agent's neighbor $j$.

\textbf{Reward function.} We design the reward to guide the agent towards the optimal policy for our desired QoS, i.e., minimizing average delivery delay, while ensuring maximum number of packets get delivered successfully.

To accomplish this, we craft our reward function to include spectrum band and next node parameters, like EM characteristics of various bands (mainly, transmission range and bit rate), channel availability (no harmful interference to PUs and other nearby SUs), and queuing at each node. If an agent $i$ (at state $s_t$) is transmitting packet $p$ to next node $j$ over band $b$, with the packet destination node being $y_p$, then the reward is given by:
\begin{equation}\label{eq:pareto mle2}
\begin{split}
     R^{a_t}_{ij} =
    \eta_1 asinh(-{T}_q^{j}) + \delta.[\theta(j)-\phi(b)] -  \\ \mu.(1-\theta(j)) + \rho[\eta_2 asinh({-T}_d^{i,j}) + \ \eta_3  asinh({T}_r^{i,j,y_p})],
\end{split}
\end{equation}

where ${T}_q^{j} = \frac{n_j \times l}{\mathcal{R}_{min}}$ is the worst-case queuing delay at next node $j$. Here $n_j$ is the number of packets in the queue of $j$, $\mathcal{R}_{min}$ is the minimum data rate among all bands and $l$ is the packet size. $T_d^{(i,j)} = \frac{l}{\mathcal{R}_b}$ is the transmission delay in transmitting the packet to next node $j$, where $\mathcal{R}_b$ is the data rate for band $b$. $T_r^{(i,j,y_p)} = (d_{iy_p} - d_{jy_p})$ is the difference in distance between agent $i$ to destination $y_p$ and next node $j$ to $y_p$. A positive value denotes that the packet is routed towards the destination. Since we want to minimize the delay, we have a negative sign associated with it in the reward. \emph{asinh} function normalizes the undertaken value to a common scale.  

To differentiate between the destination node and a typical relay node for a packet $p$, we assign a higher reward if the next relay node selected is a destination node. Otherwise we penalize the action. Parameters $\delta$ and $\mu$ are used for this purpose. $\theta(j) \in [0, 1]$ denotes whether the chosen next node $j$ is the destination or not. $\phi(b) \in [0, 1]$ represents the channel unavailability where 1 denotes that a common channel over selected band is unavailable. This, along with the transmission delay, helps in selecting alternative bands. The queuing delay, the distance difference parameters in conjunction with $\delta$ and $\mu$ aid in next hop node selection.  $\eta_1$, $\eta_2$, $\eta_3$ are hyperparameters which can be varied to suit performance needs.

Since d-DSA networks change over time (mainly due to PU channel usage activities), the state transition probabilities and reward functions cannot be modeled. Thus, we utilize a model-free RL technique, called Q-learning, to find the optimal cross-layer band selection and routing policy.

According to this approach, each d-DSA node (or agent) in the network maintains a Q-table which is used to select an action at each state. The states are represented as rows and the actions as columns in the Q-table. The values in the Q-table represent the quality of a particular action (Q-value) for a particular state. The quality of the action is represented by the long term reward gained by executing that action. In our RL framework, at time stamp $t$, each d-DSA node observes the state $s_t$, and selects an action $a_t$  from its action space by looking up the Q-table. The reward $r_{t+1}$ obtained provides feedback on how well the action performed. The Q-value function maintained in the Q-table can be approximated to the optimal value by updating and iterating according to 
\begin{equation}
\label{eqn:qlearnupdate}
    Q(s_t, a_t) \leftarrow (1 - \alpha) Q(s_t, a_t) + \alpha [r_t + \gamma max_{a'} Q(s_{t+1}, a')]
\end{equation}

where $\alpha \in (0, 1]$ is the learning rate, which models the rate of updating Q-values. $\gamma \in [0,1]$ is the discount factor, which is used to discount the future rewards such that the recent actions affect the current value more than the future ones.  

Algorithm \ref{P0_algo} details the proposed RL-based BARD algorithm. In Line 1-5, we compute the action space for each d-DSA node $i \in U$. At each simulation time stamp $\tau$, time is divided into further steps where each step denotes a Q-table update for the packet in consideration. Then, each  d-DSA follows an $\epsilon$ - greedy policy (line 11 - 16), according to which the d-DSA node \textit{explores} the action space and \textit{exploits} what it has learnt with probability $\epsilon$ and ($1 - \epsilon)$ respectively. During the initial phase, BARD \emph{explores} more and later, it \emph{exploits} what it has learnt. This is achieved by employing a decaying $\epsilon$ as the exploration factor. In line 17, it computes the required parameters for reward function, i.e.,  $T_q^j$, $T_d^{ij}$, and $T^{i,j,y_p}_r$. Next, we utilize the selected action, say $a_t<j,b>$ and send the packet to next node $j$ over band $b$, if the time duration permits and common channel is available (See lines 18 - 22). Finally, whatever reward is obtained after executing the action, is used in (\ref{eqn:qlearnupdate}) to update the Q-table (line 26). This happens at each node for all the message packets. With more packets, the algorithm moves nearer to the optimal policy. 


   \begin{algorithm}
   \textbf{Input:} d-DSA node $i$, Q, $\alpha$, $\epsilon$,$\epsilon_{min}$, $\epsilon_{decay}$, $\eta_1, \eta_2, \eta_3, \delta, \mu, \gamma$ \\
	\textbf{Output:} $Q(s,a)$, $s \in S$, $a \in A :$ Q-table matrix 
	\begin{algorithmic}[1]
	\State Initialize action-space, $A_i = \phi$
	\For{each node $j \in U$}
	\For{each band $b \in \mathcal{B}$}
	\If{$ d_{ij} \leq \Delta_b$}
	\State Include $a<j, b> \in A_i$ 
	\EndIf
	\EndFor
	\EndFor
	\State Exploration parameter, $\epsilon = 1$
    \For{$\tau = 0  \to T$}
        \State $t = \tau$
        \State //$\tau_{\Delta}$ represents the time interval between $\tau$ and $(\tau + 1)$
        \State rem-time = $\tau_{\Delta}$
        \For{each packet $p \in$ d-DSA node $i$'s queue}
        \State Observe state $s_{t}<y_p, m_q, E_i, E'_i>$
        \State rand-number = random(0,1)
        \If{rand-number $\leq \max(\epsilon, \epsilon_{min})$}
        \State Choose a random action $a_t<j,b>$ from $A_i$ 
        \Else
        \State $a_{t}<j,b> = \arg\max_{a' \in A_i} Q(s_{t}, a'))$
        \EndIf
        \State $\epsilon = \epsilon \times \epsilon_{decay}$
        \State Compute $T_q^j$, $T_d^{ij}$, and $T^{i,j,y_p}_r$ (See Eq. \ref{eq:pareto mle2})
        \If{$T_d^{ij} < $ rem-time}
        \If{common channel is available}
        \State Send packet to next node $j$ over band $b$ 
        \State rem-time = rem-time $- T_d^{ij}$  
        \State $\theta(j)$ = 1
        \Else 
        \State $\phi(b)$ = 1
        \EndIf
        \EndIf
        \State Go to next state $s_{t+1}$, compute reward $r_{t}$ (Eq. \ref{eq:pareto mle2})
        \State$Q(s_{t}, a_{t}) \leftarrow (1 - \alpha) Q(s_{t}, a_{t}) + \alpha [r_{t} + \gamma \times max_{a} Q(s_{t +1}, a)]$
        \State t = t + 1
        \EndFor
        \EndFor 
	\end{algorithmic}  
	\caption{Proposed BARD algorithm (at d-DSA node $i$)}
	\label{P0_algo}  
\end{algorithm}

\section{Performance Evaluation} \label{sec:results}
In this section we evaluate the performance of our BARD algorithm against baseline dDSAaR algorithm and single band variants (both described later in this section) in terms of following performance metrics, typically used in wireless networks -- (1) \textit{Message Delivery ratio}: Fraction of total messages successfully delivered to the destination nodes divided by the total messages generated at the source nodes., and (2) \textit{Network Latency}: The average delay incurred (transmission delay, propagation delay, and queuing delay) in delivering all messages from source to the destination.  

For extensive analysis, each of these performance metrics is evaluated against important parameters -- \emph{simulation time}, \emph{number of primary users} and \emph{number of secondary users}. To further highlight the benefits of BARD algorithm, we present the \textit{band usage analysis} for specific scenarios where all PUs are concentrated in a single band of interest.

\begin{figure*}[h!]
\centering
\begin{subfigure}[t]{0.45\linewidth}
    \includegraphics[width=\linewidth]{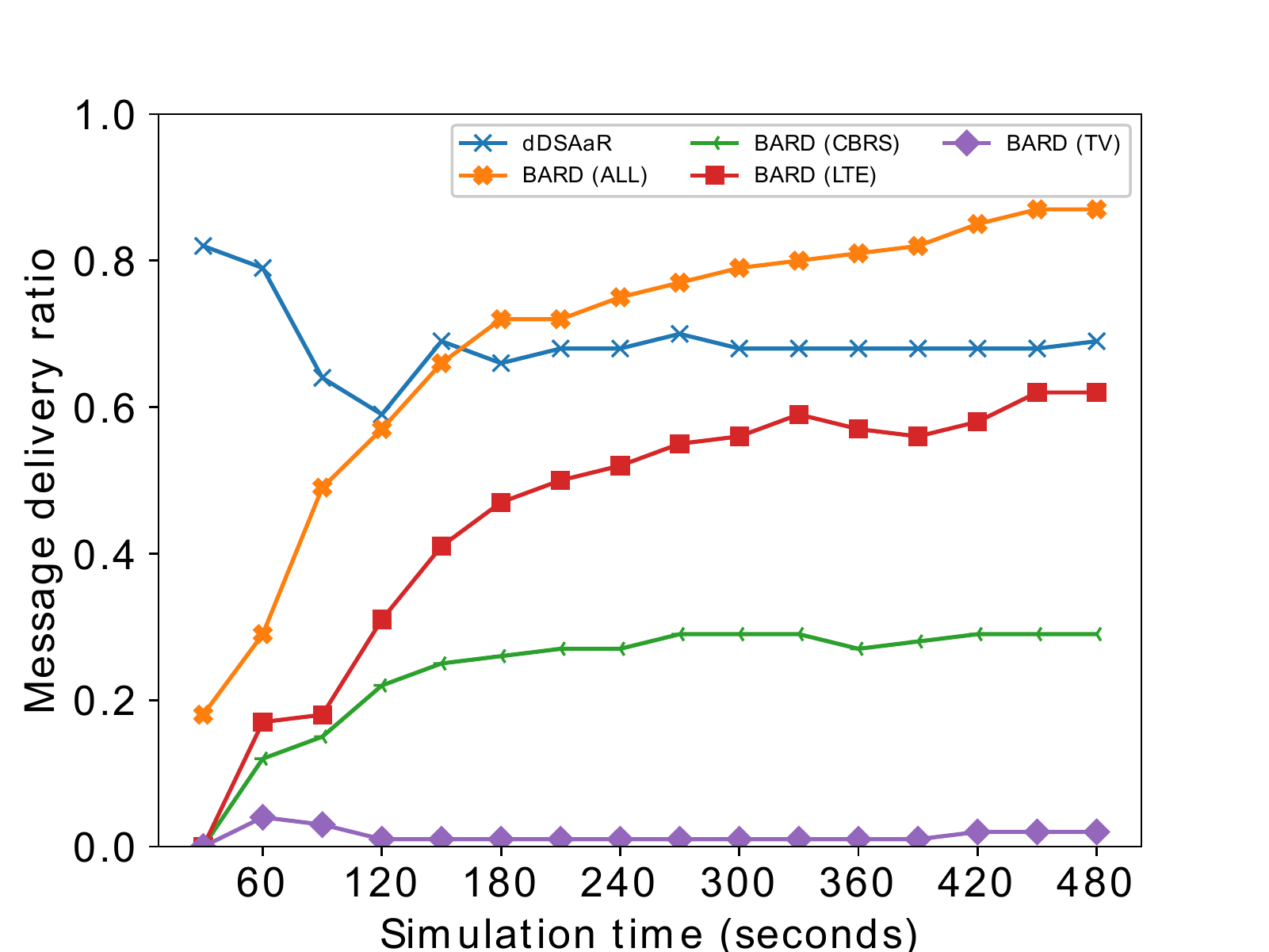}
    \caption{Simulation time}
    \label{fig:MDR_time}
\end{subfigure}%
\begin{subfigure}[t]{0.45\linewidth}
    \includegraphics[width=\linewidth]{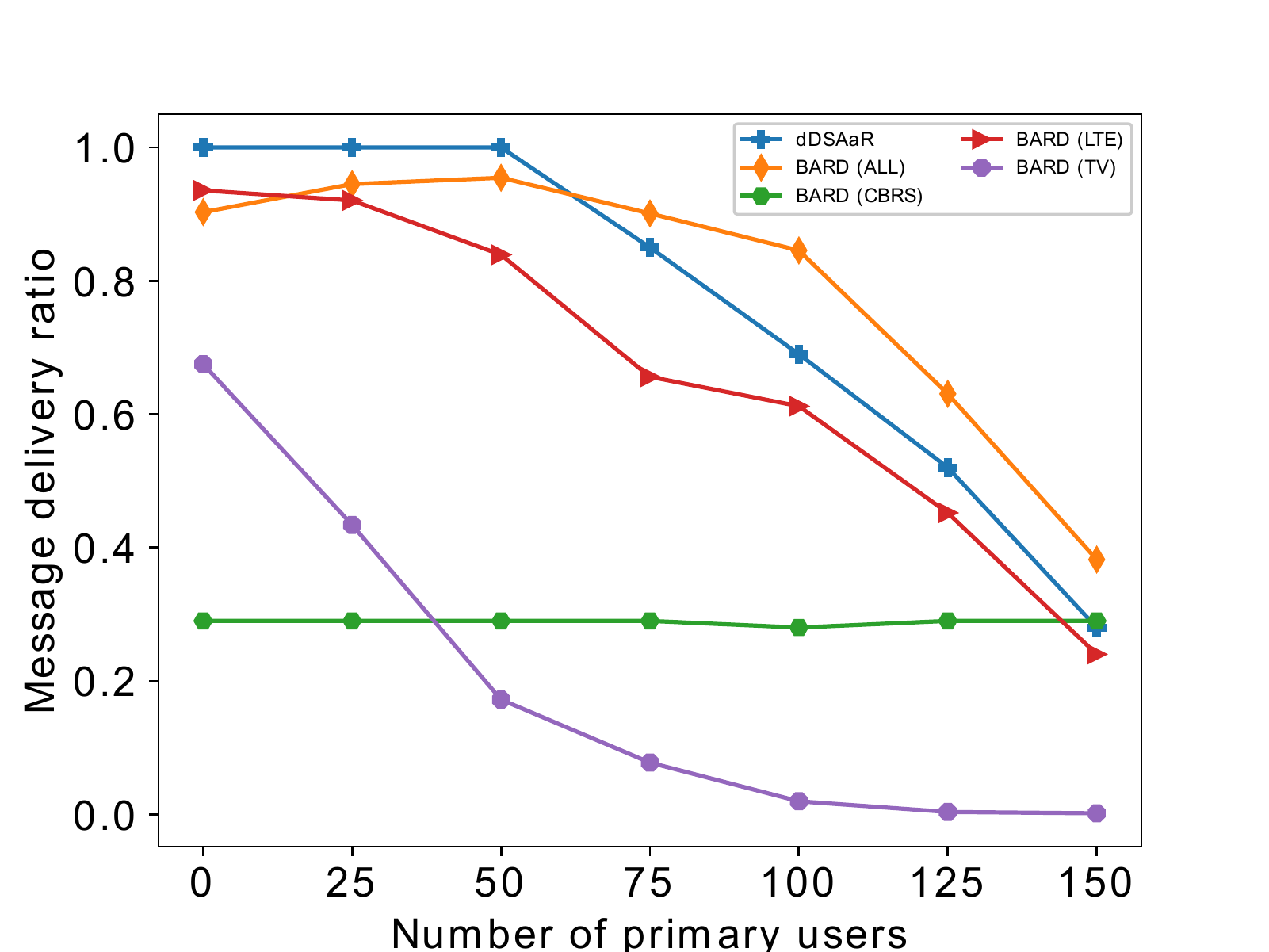}
    \caption{Number of PUs}
    \label{fig:MDR_PU}
\end{subfigure}
\begin{subfigure}[t]{0.45\linewidth}
    \includegraphics[width=\linewidth]{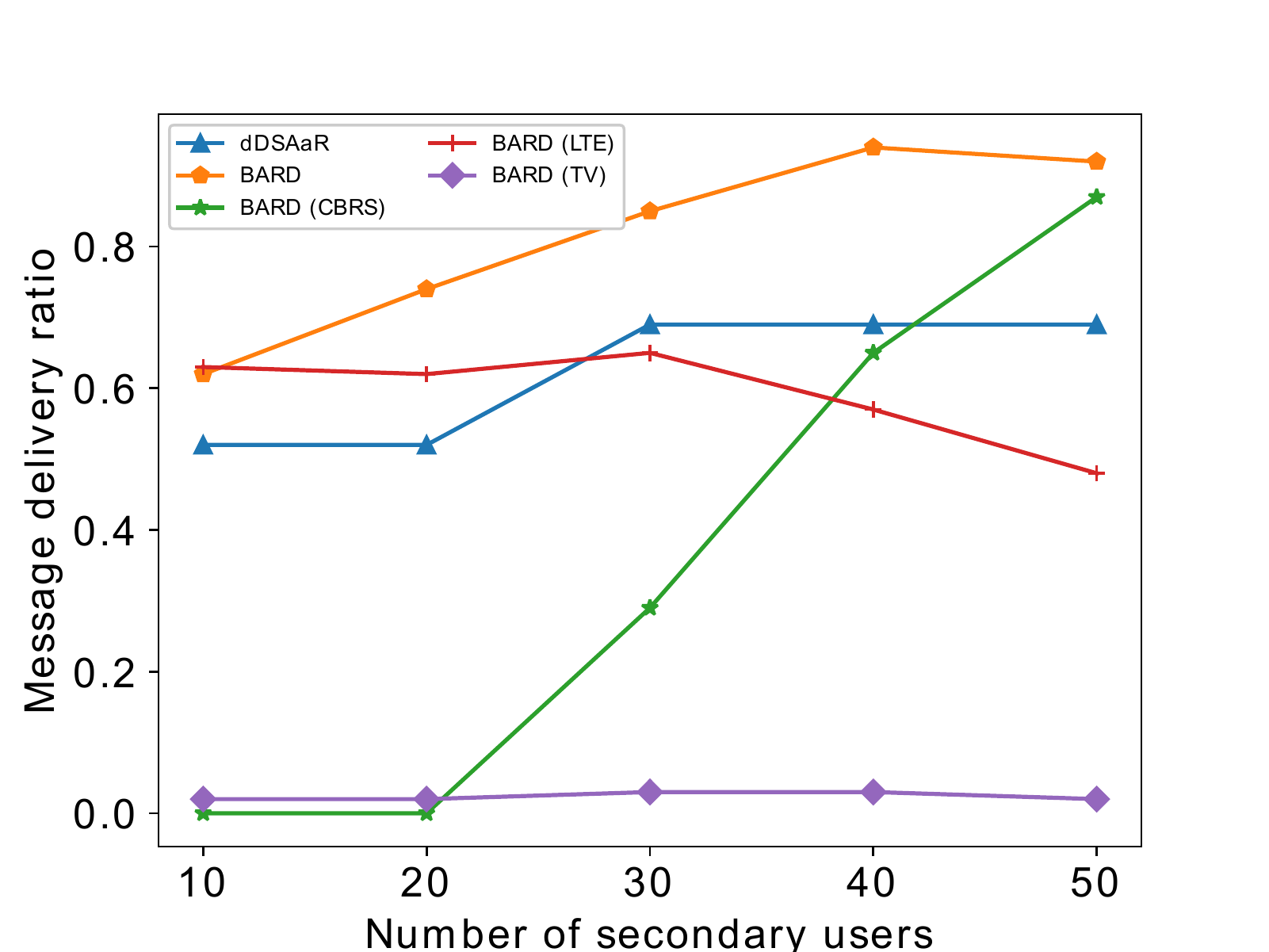}
    \caption{Number of SUs}
    \label{fig:MDR_SU}
\end{subfigure}
\begin{subfigure}[t]{0.45\linewidth}
    \includegraphics[width=\linewidth]{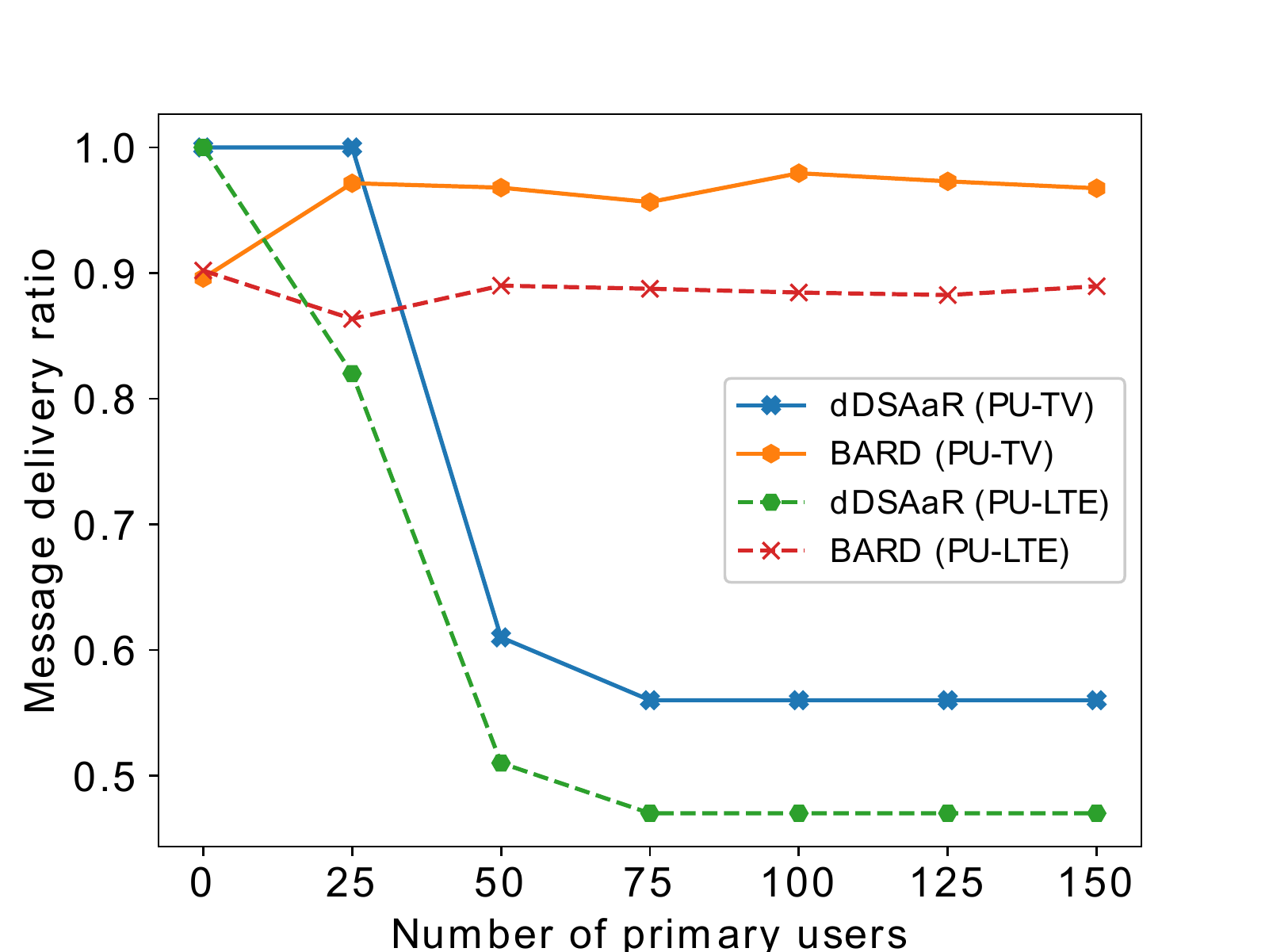}
    \caption{All PUs in one band}
    \label{fig:MDR_spec_case}
\end{subfigure}
\vspace{-0.05in}
\caption{Message Delivery Ratio: (a) Simulation time, (b) Number of PUs, and (c) Number of SUs (d-DSA nodes), and (d) All PUs concentrated in one primary band (TV or LTE)} \label{fig:MDR}
\vspace{-0.2in}
\end{figure*}

\begin{figure*}
\centering
\begin{subfigure}[t]{0.45\linewidth}
    \includegraphics[width=\linewidth]{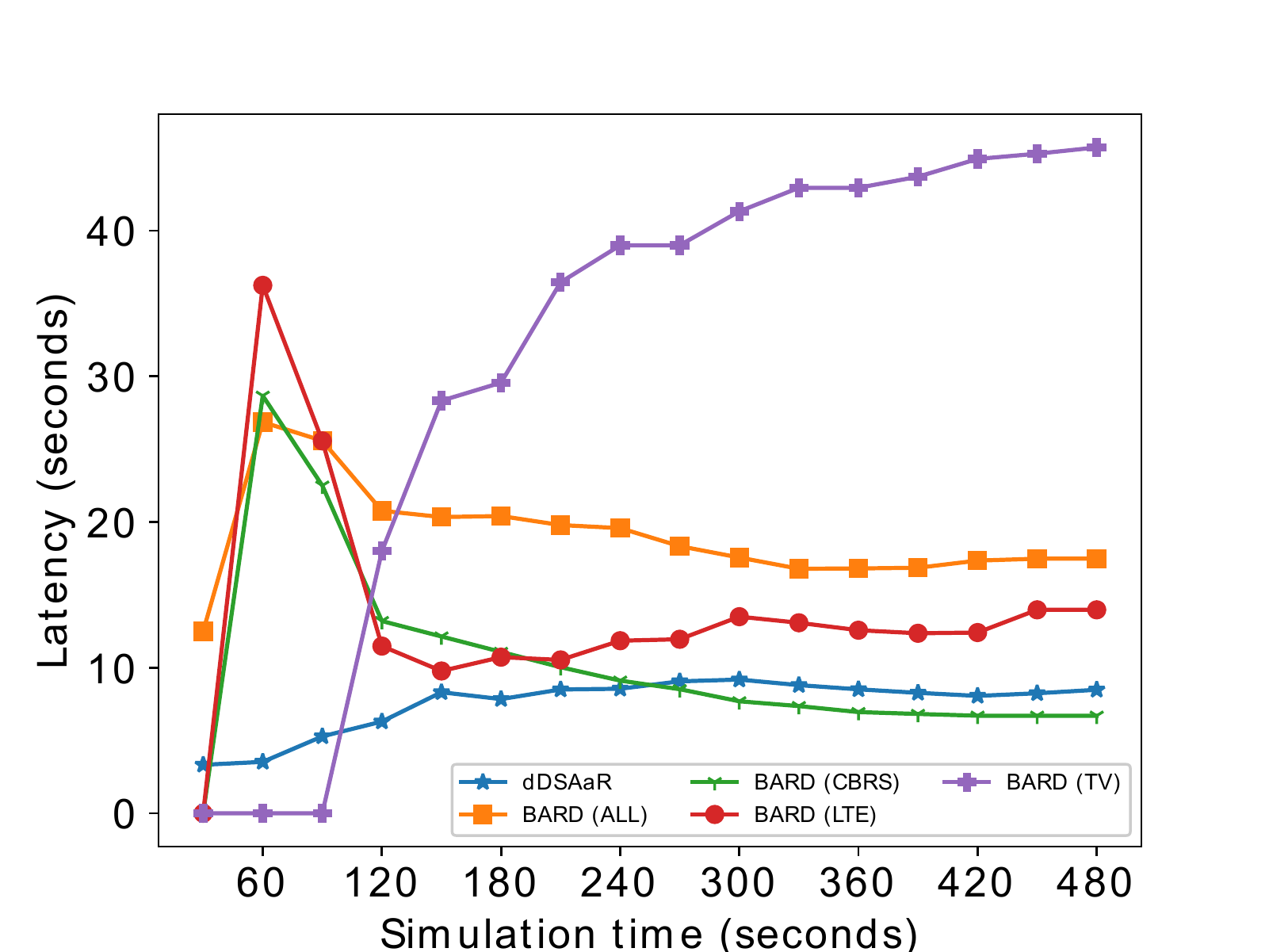}
    \caption{Simulation Time}
    \label{fig:latency_time}
\end{subfigure}%
\begin{subfigure}[t]{0.45\linewidth}
    \includegraphics[width=\linewidth]{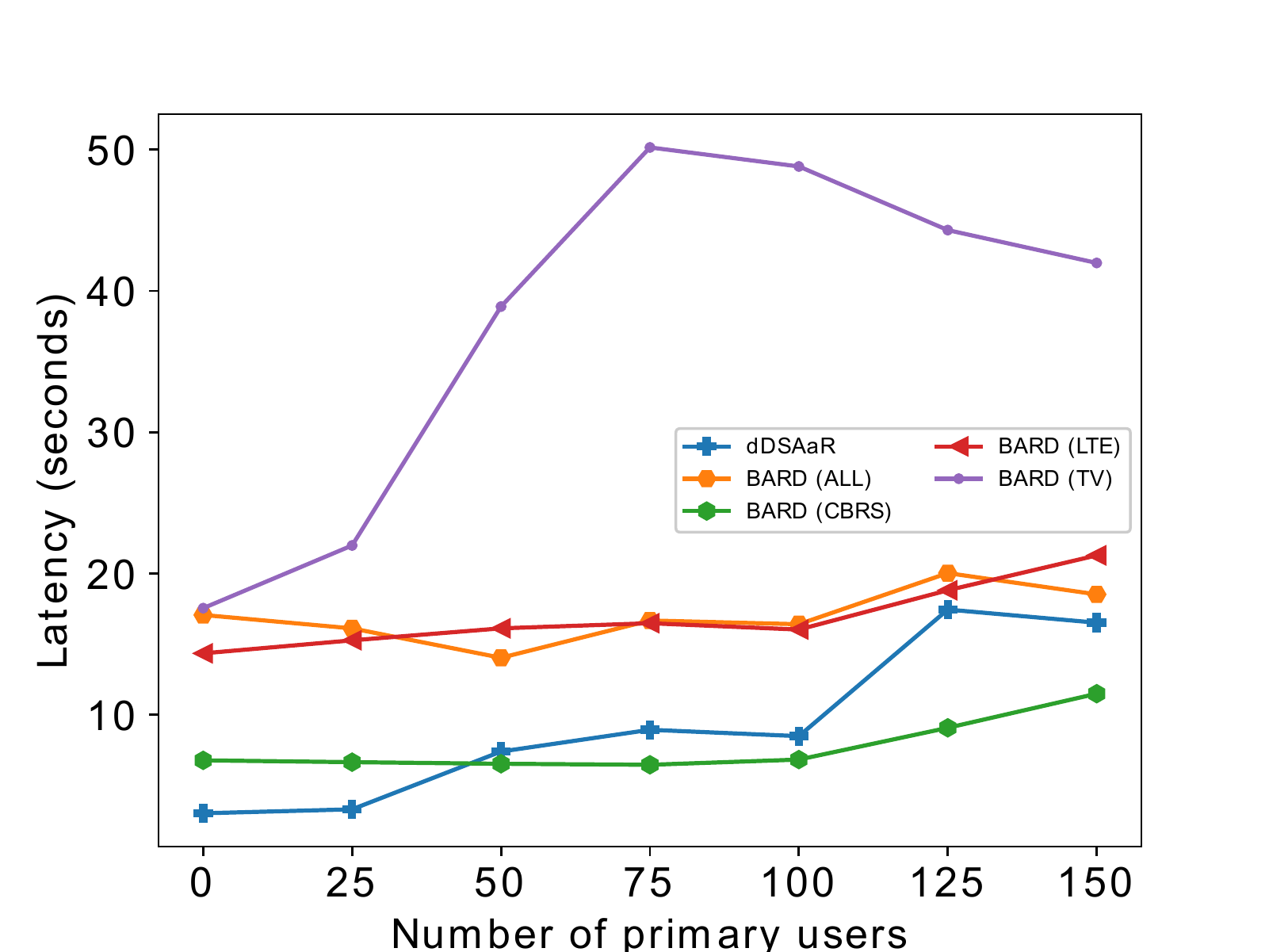}
    \caption{Number of PUs}
    \label{fig:latency_PU}
\end{subfigure}
\begin{subfigure}[t]{0.45\linewidth}
    \includegraphics[width=\linewidth]{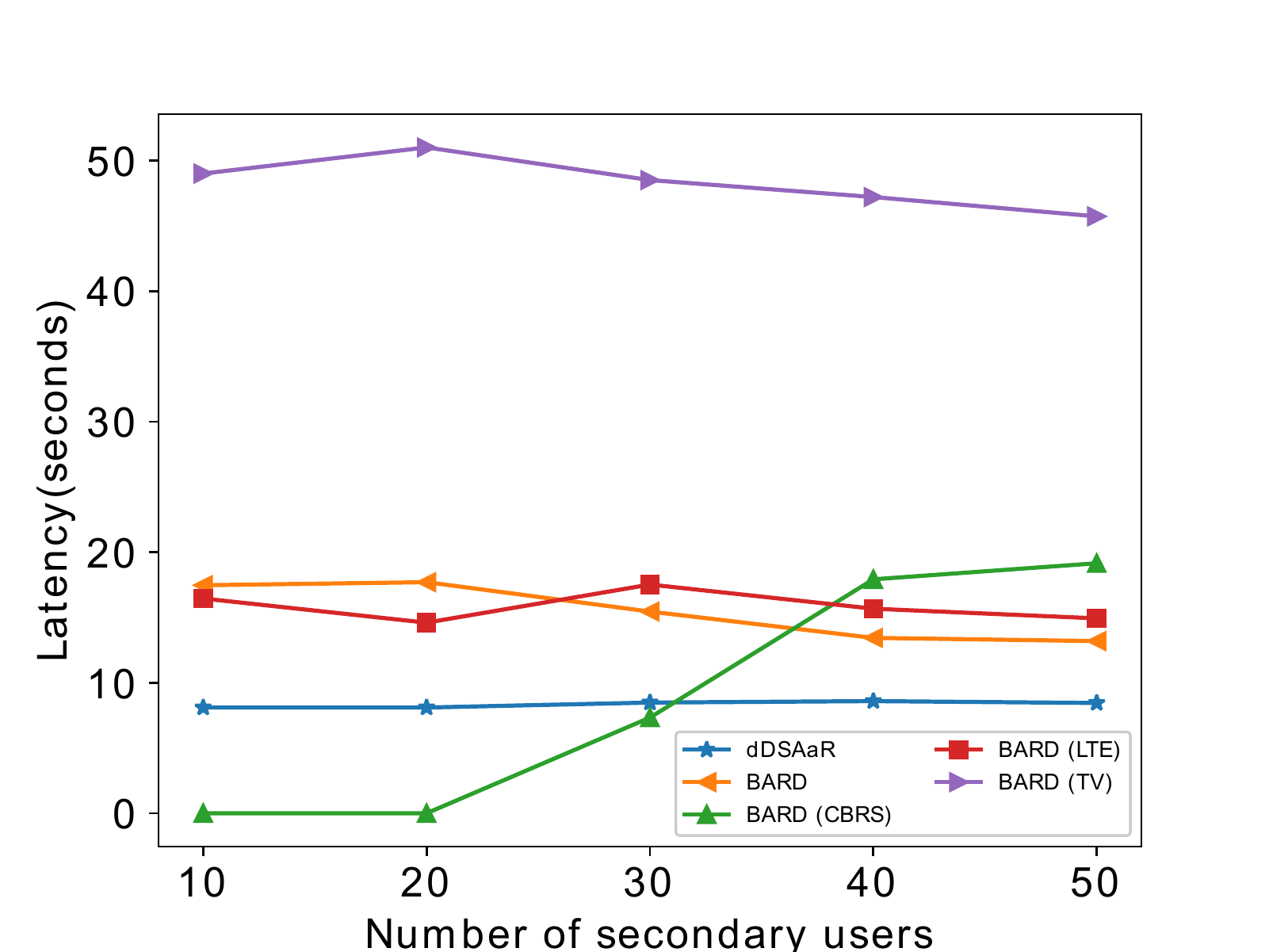}
    \caption{Numbers of SUs}
    \label{fig:latency_SU}
\end{subfigure}
\begin{subfigure}[t]{0.45\linewidth}
    \includegraphics[width=\linewidth]{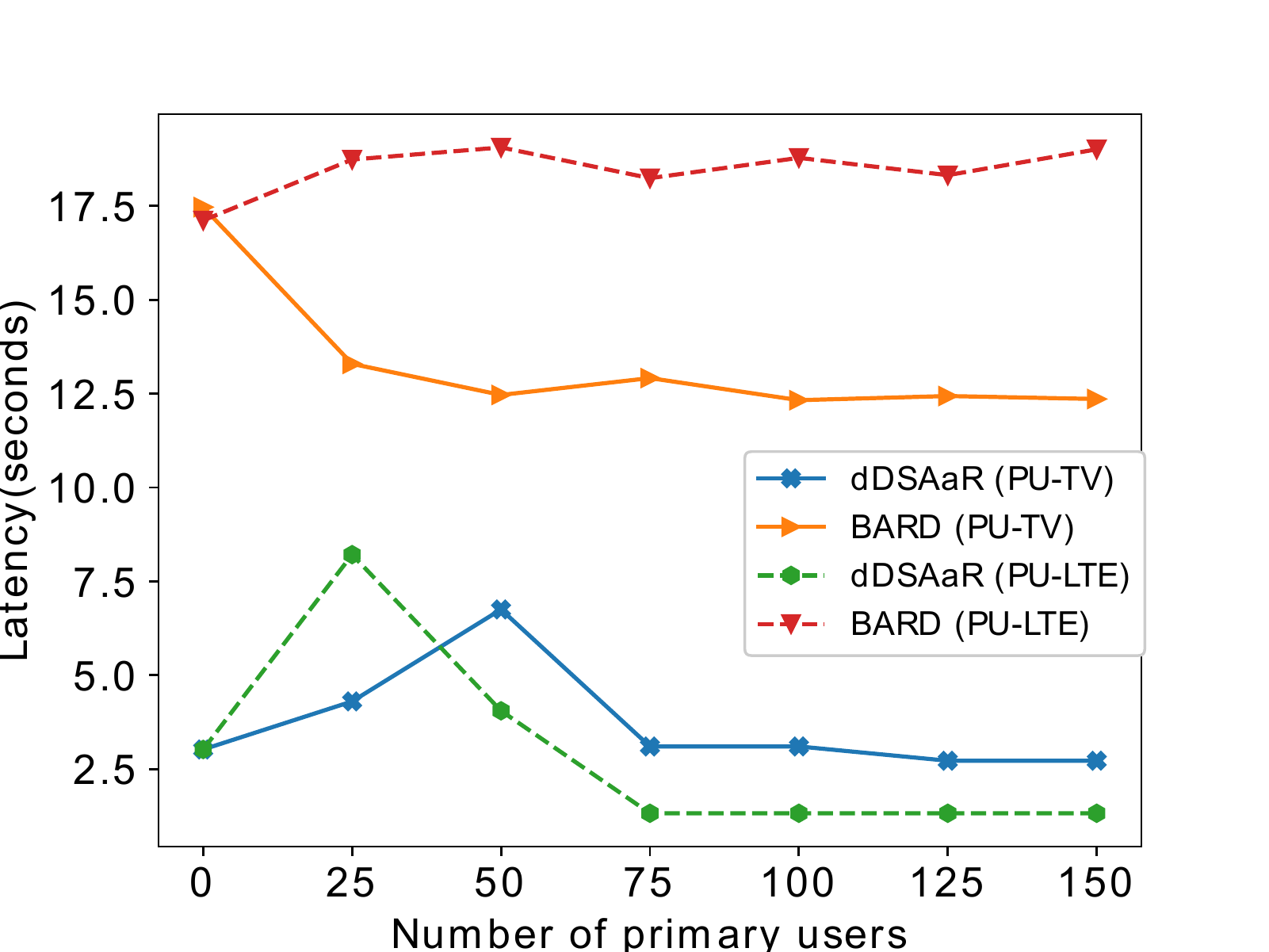}
    \caption{Numbers of SUs}
    \label{fig:latency_spec_case}
\end{subfigure}
\vspace{-0.05in}
\caption{Network Latency: (a) Simulation time, (b) Number of PUs, (c) Number of SUs (d-DSA nodes), and (d) All PUs concentrated in one primary band (TV or LTE)} \label{fig:latency}
\vspace{-0.23in}
\end{figure*}

\textbf{Simulation Setting.}
We consider a simulation setting with an area of $4$ $km^2$ where d-DSA nodes and PUs are deployed randomly. Both PUs and SUs locations are fixed and stationary, though link existence (over a certain band) may vary over time. Unless otherwise stated, all experiments are performed with $150$ PUs and $30$ SUs (or d-DSA nodes), out of which, 3 are source nodes, 3  are destination nodes and rest $24$ act as relay nodes. The total simulation time is $480$ seconds. 

\textit{PU and channel model.}  The ON/OFF time for a PU in a certain channel within a band is chosen randomly from the interval [$1-4$] seconds. The channel availability within each band is subject to change over time as PU's activity changes. We consider that each spectrum band has $6$ channels. 

\textit{Message traffic model.} We consider the message traffic model with bursty generation of messages at each source node. The length of each burst varies between [$5-15$] messages and the inter-arrival time between the bursts is exponentially distributed with a mean of $15$ seconds. The size of each generated message is randomly chosen from [$5, 20, 40, 60$] Mb. The messages are transmitted as packets with each packet having a size of $5$ Mb (Mega bit) and a TTL of $60$ seconds. 

\textit{Transmit power and bit rate.} The transmit power for various spectrum bands are considered as follows -- 4 W for TV and LTE, 1 W for ISM and 10 W for CBRS band (as allowed by FCC). The transmission range (using close-in (CI) free space reference distance model~\cite{sun2016propagation}) and effective bit rate (from Shannon Hartley theorem) were computed by considering a path loss factor of $2.8$, received power threshold of $-95$ dBm and Signal-to-Noise Ratio (SNR) of $-100$ dBm.  


\textit{Other parameters.} The values of RL parameters are as follows: learning rate $\alpha=0.2$, discount factor $\gamma=0.6$, exploration factor $\epsilon = 1$, decay $\epsilon_{decay}=0.95$, minimum exploration factor $\epsilon_{min} = 0.1$, hyperparameters $\eta_1 = \eta_2 = \eta_3 = 2$, destination reward $\delta = 10$, and penalty $\mu = 2.5$. Since learning takes place in an online fashion, there is some variation in results to be expected. To reduce this, we perform simulations over $20$ rounds and showcase the averaged results.

\textbf{Comparison approaches.} \label{subsec:comparison_approaches}
We evaluate BARD algorithm against the following comparison approaches.
\paragraph{dDSAaR protocol~\cite{shah2019x}}
dDSAaR is a centralized routing protocol originally proposed for time-varying yet sufficiently predictable d-DSA networks. It works in two discrete rounds - \textit{bootstrap} and \textit{operation}. In bootstrap round, dDSAaR utilizes the historic mobility information to construct a unified Space-Time-Band (STB) graph that captures the communication opportunities over space, time, and diverse bands. 
See Section V.A. in \cite{shah2019x} for details on STB graph. Based on the STB graph, dDSAaR uses an optimal least delay cost (LDC) path algorithm to compute the minimum delay routes for each source-destination node pair in the d-DSA network, which are then utilized in the subsequent operation round, where the node's mobility routes may have changed. 

Notice that since we consider fixed location of all d-DSA nodes, the potential communication links over various bands are constant and do not change over time. 
This means that the optimal routes computed in the bootstrap round will remain the same in the operation round, thus, potentially better network performance.  
However, it is worthwhile mentioning that dDSAaR, unlike our proposed BARD algorithm, does not consider the real-time traffic or channel usage related events in the network, e.g., queuing delay at relay nodes, and channel usage in a certain band due to PU activities etc, while will negatively impact the performance of dDSAaR protocol. 


\paragraph{Single-band variants} For the sake of completeness, we compare our BARD algorithm that harnesses all spectrum bands in the d-DSA networks against its single-band variants. It represents utilizing BARD algorithm for the conventional (single-band) DSA networks, and the functionality of BARD is reduced to only choosing next hop node while ensuring non-interference with PUs over the prespecified individual primary band. For the clarity of plots, we do not show the plots for dDSAaR single-band variants. However dDSAaR and BARD single-band variants show similar comparative behavior with that of dDSAaR and BARD for all spectrum bands.

\textbf{Simulation Results.} Now we discuss the experimental results of BARD and other comparison approaches.
\subsubsection{MDR Analysis} 
From Fig. \ref{fig:MDR_time}, we observe that BARD significantly outperforms the dDSAaR routing protocol (after $120$ s) and its single band variants in terms of MDR, for varying simulation times. While the performance of BARD is not ideal in the beginning, it can be seen that it quickly catches up in terms of overall MDR. The initial phase of low performance which exists for all variations of BARD is due to the protocol learning the network and RF characteristics. 

The MDR for dDSAaR is quite high for lower number of primary users (see Fig. \ref{fig:MDR_PU}), however, it gradually declines as the PU interference in utilized bands increases. The TV band which has the maximum number of links suffers greatly from PU and SU interference. Its higher range leads to channel unavailability issues with many users vying for limited number of channels in the band. This, coupled with TV bands' lower data rate, leads to queuing and dropping of packets due to their TTL being exceeded. The CBRS band has a constant performance for varying time and primary users mainly due to the absence of paths over CBRS band between each source-destination pairs. Both the graphs highlight the benefit of using multiple bands instead of a single band. Greater number of SUs maintains the MDR but it was observed that it gives rise to greater number of hops due to increased node density and therefore more routes to transmit the packet. 

For dDSAaR, the number of routes in the network does not matter much as it uses the pre-computed path. So though the MDR rises for more than 20 nodes, it quickly saturates since the new nodes added do not give a better LDC path. But BARD is able to take full advantage of the new links added to the network till the MDR is close to 1. The dip in performance of LTE is due to increased interference from the secondary users. There is a steep increase in MDR for CBRS as the number of links increase due to the very high data rate it offers.  
It is worthwhile to note that even though BARD suffers from higher secondary user interference in conjunction with PUs than dDSAaR, it is still able to outperform dDSAaR and single band variants due to better band and route selection.

\subsubsection{Latency Analysis}
Fig \ref{fig:latency_time} shows that the incurred latency is high in the initial stages for all variants of BARD ($60$ s). This is due to the training phase which takes place in an online manner but stabilizes in the later stages as the protocol reaches the optimal policy. As seen in Fig \ref{fig:latency_PU}, the average latency in dDSAaR increases due to PU interference which implies channel unavailability and queuing at the nodes. The number of primary users does very little to affect the latency of BARD since it is capable of finding alternative routes even in the relatively dense scenario we have considered. The lesser latency of dDSAaR can also be attributed to the fact that lesser number of packets are being delivered and we only consider the contribution of delivered packets.
Fig \ref{fig:latency_SU} shows that the latency is decreasing for BARD. The decrease is due to the increase in the available routes towards the destination. But dDSAaR maintains a constant latency since the increased number of SUs contribute little to the LDC path computation. TV band suffers from the worst latency mainly due to very high interference from both PUs and other SUs in the area.

\begin{table} [h!]
\vspace{-0.07in}
\caption{{\upshape Band usage (PUs concentrated in a single band)}}\label{bandUsage} 
\centering 
\begin{tabular}{|p{0.9cm}|p{0.35cm} | p{0.35cm}|p{0.35cm}|p{0.4cm}|p{0.5cm} | p{0.45cm}|p{0.35cm}|p{0.45cm}|p{0.5cm}|} 

\hline 
Protocol & PUs & \multicolumn{4}{c|}{All PUs in TV band}
& \multicolumn{4}{c|}{All PUs in LTE band} \\

\cline{3-10}
&& TV & ISM & LTE & CBRS &  TV & ISM & LTE & CBRS \\  
\hline 

\multirow{4}{*}{dDSAaR} & $0$ & $15.2$ & $0$ & $53.2$  & $31.6$ & $15.1$ & $0$ & $53.2$ & $31.6$  \\
& $50$ & $1.2$ & $0$ & $62.0$  & $36.8$ & $56.5$ & $0$  &  $13.4$ & $30.1$ \\
& $100$ & $0.1$ & $0$ & $62.7$  & $37.2$ & $67.8$ & $0$ & $0$ & $32.2$ \\
& $150$ &  $0.05$ & $0$ & $62.7$  & $37.2$ & $67.8$ & $0$ & $0$ & $32.2$  \\ \hline
 
\multirow{4}{*}{BARD} & $0$ & $39.5$ & $6.5$ & $38.1$  & $15.9$ & $39.8$ & $6.2$ & $38.8$ & $15.2$  \\
& $50$ & $4.9$ & $13.2$ & $43.9$  & $38.0$ & $61.7$ & $8.7$  &  $2.6$ & $26.8$ \\
& $100$ & $1$ & $11.5$ & $40.6$  & $46.9$ & $70.6$ & $10$   & $0.04$ & $19.3$ \\
 &$150$ &  $0.2$ & $30$ & $16.9$  & $52.9$ & $72.8$ & $6.9$ & $0.02$ & $20.2$  \\ \hline
\end{tabular}
\end{table}

\subsubsection{Band Usage Analysis}
The band usage analysis has been carried out for the scenario where all primary users are concentrated in a single band. This is performed to model a highly interfered band and to show how dDSAaR suffers even if the other bands were interference free. This is evident from Fig. \ref{fig:MDR_spec_case} where BARD outperforms dDSAaR handsomely for higher number of PUs. Table \ref{bandUsage} provides an overview of the band usage statistics of both the routing protocols. It can be seen that BARD is able to make efficient use of alternative bands when it learns the channel unavailability in the affected band. It even makes use of the ISM band which was not used before and is used so that the load on LTE and CBRS does not increase and thus compromise performance. The performance of dDSAaR is constant for higher PUs. This only means that the affected band is fully unavailable and only packets whose path don't include the affected band are transmitted successfully. This is reflected in the latency plot as well. The table and plots both indicate that LTE was a major contributor than TV and without the LTE band, the MDR is lower while the delay is higher for both BARD and dDSAaR. Fig. \ref{fig:latency_spec_case} shows that BARD has higher latency than dDSAaR. This is to be expected since not all the messages are being delivered and the ones that are being delivered also do not suffer any interference from PUs and negligible interference from SUs. 



\section{Conclusions} \label{sec:conclusions}
This paper proposed a decentralized, online multi-agent reinforcement learning based cross-layer band selection and routing (BARD) algorithm for diverse band-aware DSA (d-DSA) networks. Specifically, BARD utilized whitespaces in multiple spectrum bands and harnessed their unique EM properties to maximize QoS requirements of heterogeneous messages; while ensuring no harmful interference to the primary users in the utilized band. Our experiments showcased that BARD notably outperforms the baseline dDSAaR protocol in terms of message delivery ratio, at a slightly higher network latency, against all parameters. Nevertheless, BARD greatly outperforms its single-band variants in both the metrics. 

\bibliography{main.bib}

\begin{thebibliography}{10}

\bibitem{NSF-SII}
NSF SII.
\newblock https://www.nsf.gov/pubs/2020/nsf20557/nsf20557.htm.

\bibitem{NSF-SWIFT}
NSF SWIFT.
\newblock https://www.nsf.gov/pubs/2020/nsf20537/nsf20537.htm.

\bibitem{presidential-memorandum}
US~Whitehouse.
\newblock https://www.whitehouse.gov/presidential-actions/ \\
  presidential-memorandum-developing-sustainable-spectrum-strategy-\\americas-future/.

\bibitem{jagannath2018design}
J.~Jagannath, S.~Furman, T.~Melodia, and A.~Drozd.
\newblock Design and experimental evaluation of a cross-layer deadline-based
  joint routing and spectrum allocation algorithm.
\newblock {\em IEEE Trans. on Mobile Comp.}, 2018.

\bibitem{arslan2007cognitive}
H.~Arslan and J.~Mitola~III.
\newblock Cognitive radio, software-defined radio, and adaptive wireless
  systems.
\newblock {\em Wireless Communications and Mobile Computing}, 7(9):1033--1035,
  2007.

\bibitem{tang2016joint}
F.~Tang and J.~Li.
\newblock Joint rate adaptation, channel assignment and routing to maximize
  social welfare in multi-hop cognitive radio networks.
\newblock {\em IEEE Trans. on Wireless Communications}, 16(4):2097--2110, 2016.

\bibitem{chowdhury2011crp}
K.~R Chowdhury and I.~F Akyildiz.
\newblock Crp: A routing protocol for cognitive radio ad hoc networks.
\newblock {\em IEEE Journal on Selected Areas in Communications},
  29(4):794--804, 2011.

\bibitem{shah2017designing}
V.~K Shah, S.~Bhattacharjee, S.~Silvestri, and S.~K Das.
\newblock Designing sustainable smart connected communities using dynamic
  spectrum access via band selection.
\newblock In {\em ACM International Conference on Systems for Energy-Efficient
  Built Environments (BuildSys)}, pages 1--10, 2017.

\bibitem{shah2019diverse}
V.~K Shah.
\newblock A diverse band-aware dynamic spectrum access architecture for
  connectivity in rural communities.
\newblock {\em Ph.D. Thesis}, 2019.

\bibitem{shah2019x}
V.~K Shah, S.~Silvestri, B.~Luciano, and S.~K Das.
\newblock X-chant: A diverse dsa based architecture for next-generation
  challenged networks.
\newblock In {\em IEEE Conference on Computer Communications (INFOCOM)}, pages
  586--594, 2019.

\bibitem{elrhareg2019routing}
H.~Elrhareg, M.~Ridouani, and A.~Hayar.
\newblock Routing protocols on cognitive radio networks: Survey.
\newblock In {\em 2019 IEEE International Smart Cities Conference (ISC2)},
  pages 296--302. IEEE, 2019.

\bibitem{liurl2020}
L.~{Li}, L.~{Liu}, J.~{Bai}, H.~{Chang}, H.~{Chen}, J.~D. {Ashdown},
  J.~{Zhang}, and Y.~{Yi}.
\newblock Accelerating model free reinforcement learning with imperfect model
  knowledge in dynamic spectrum access.
\newblock {\em IEEE Internet of Things Journal}, pages 1--1, 2020.

\bibitem{rlCRHasan}
H.~Al-Rawi, K.-L. Yau, H.~Mohamad, N.~Ramli, and W.~Hashim.
\newblock Reinforcement learning for routing in cognitive radio ad hoc
  networks.
\newblock {\em TheScientificWorldJournal}, 2014:960584, 07 2014.

\bibitem{rlbandchannel2019}
S.~Jang, C.-H Han, K.~Lee, and S.-J. Yoo.
\newblock Reinforcement learning-based dynamic band and channel selection in
  cognitive radio ad-hoc networks.
\newblock {\em EURASIP Journal on Wireless Comms. and Networking}, 2019, 12
  2019.

\bibitem{rlroutingsurvey}
Z.~{Mammeri}.
\newblock Reinforcement learning based routing in networks: Review and
  classification of approaches.
\newblock {\em IEEE Access}, 7:55916--55950, 2019.

\bibitem{DeepRLMultiDSA}
O.~Naparstek and K.~Cohen.
\newblock Deep multi-user reinforcement learning for distributed dynamic
  spectrum access.
\newblock {\em Trans. Wireless. Comm.}, 18(1):310–323, January 2019.

\bibitem{shah2018designing}
V.~K Shah, S.~Bhattacharjee, S.~Silvestri, and S.~K Das.
\newblock Designing green communication systems for smart and connected
  communities via dynamic spectrum access.
\newblock {\em ACM Transactions on Sensor Networks (TOSN)}, 14(3-4):1--32,
  2018.

\bibitem{shah2020diverse}
V.~K Shah, B.~Luciano, S.~Silvestri, S.~Bhattacharjee, and S.~K Das.
\newblock A diverse band-aware dynamic spectrum access network architecture for
  delay-tolerant smart city applications.
\newblock {\em IEEE Transactions on Network and Service Management}, 2020.

\bibitem{Distdynamicchsense}
H.~{Chang}, H.~{Song}, Y.~{Yi}, J.~{Zhang}, H.~{He}, and L.~{Liu}.
\newblock Distributive dynamic spectrum access through deep reinforcement
  learning: A reservoir computing-based approach.
\newblock {\em IEEE Internet of Things Journal}, 6(2):1938--1948, 2019.

\bibitem{zhang2018fully}
K.~Zhang, Z.~Yang, H.~Liu, T.~Zhang, and T.~Ba{\c{s}}ar.
\newblock Fully decentralized multi-agent reinforcement learning with networked
  agents.
\newblock {\em arXiv preprint arXiv:1802.08757}, 2018.

\bibitem{sun2016propagation}
S.~Sun, T.~S Rappaport, S.~Rangan, T.~A Thomas, A.~Ghosh, I.~Z Kovacs,
  I.~Rodriguez, O.~Koymen, A.~Partyka, and J.~Jarvelainen.
\newblock Propagation path loss models for 5g urban micro-and macro-cellular
  scenarios.
\newblock In {\em IEEE Vehicular Technology Conference (VTC Spring)}, pages
  1--6, 2016.

\end{thebibliography}
\bibliographystyle{unsrt}

\end{document}